\documentclass[12pt,a4paper]{article}

\usepackage{amsmath}
\usepackage{amssymb}
\usepackage{latexsym}
\usepackage{graphicx}
\usepackage{psfrag,fancyhdr,epsfig}

\begin{document}

%\begin{frontmatter}
%\begin{titlepage}

\vspace*{1cm}
\begin{center}
{\bf \Large {Angular profile of emission of non-zero spin fields
from a higher-dimensional black hole}}

\bigskip \bigskip \medskip

{\bf M.~Casals}$^{1,2}$, {\bf S.~R.~Dolan}$^3$, {\bf P.~Kanti}$^4$
and {\bf E.~Winstanley}$^5$

\bigskip
$^1$ {\it School of Mathematical Sciences, Dublin City University,
Glasnevin, Dublin 9, Ireland}

$^{2}$ {\it {CENTRA, Instituto Superior T\'ecnico, Lisbon,
Portugal}}

$^{3}$ {\it {School of Mathematical Sciences, University College
Dublin, Belfield, Dublin 4, Ireland}}

$^{4}$ {\it Division of Theoretical Physics, Department of Physics,
University of Ioannina, \\ Ioannina GR-451 10,  Greece}

$^{5}$ {\it School of Mathematics and Statistics, The University of Sheffield,
Hicks Building, Hounsfield Road, Sheffield S3 7RH, United Kingdom}

%\title{Angular profile of emission of non-zero spin fields
%from a higher-dimensional black hole}

%\author[label1,label2]{M.~Casals}

%\author[label3]{S.~R.~Dolan}

%\author[label4]{P.~Kanti}

%\author[label5]{E.~Winstanley}

%\address[label1]{School of Mathematical Sciences, Dublin City University,
%Glasnevin, Dublin 9, Ireland}

%\address[label2]{CENTRA, Instituto Superior T\'ecnico Lisbon,
%Portugal}

%\address[label3]{School of Mathematical Sciences, University College
%Dublin, Belfield, Dublin 4, Ireland}

%\address[label4]{Division of Theoretical Physics, Department of Physics,
%University of Ioannina, Ioannina GR-451 10,  Greece}

%\address[label5]{School of Mathematics and Statistics, The University of Sheffield,
%Hicks Building, Hounsfield Road, Sheffield S3 7RH, United Kingdom}

\begin{abstract}
Recent works have included the effect of rotation on simulations of
black hole events at the LHC, showing that the angular momentum of the black hole
cannot be ignored and it makes a non-trivial contribution for most of the
lifetime of the black hole.
A key consequence of the rotation of the black hole is that the Hawking radiation
is no longer isotropic, making it more difficult to infer space-time parameters
from measurements of the emitted particles.
In this letter we study the angular distribution of the Hawking emission of
non-zero spin particles with specific helicity on the brane.
We argue that the {\em {shape}} of the distribution could be used as a measure
of the angular momentum of the black hole.
\end{abstract}

\end{center}

%\end{titlepage}

%\begin{keyword}
%\end{keyword}

%\end{frontmatter}

\section{Introduction}

Over the past ten years, there has been a huge amount of research into
theories with Large Extra Dimensions \cite{ADD},
where the Standard Model matter fields are confined to a
four-dimensional hyper-surface (or {\emph {brane}}), with only
gravitational degrees of freedom
permitted to propagate in the higher-dimensional bulk.

One consequence of these theories is that the fundamental
energy scale of quantum gravity, $M_{\ast }$, is related to the four-dimensional
Planck energy $M_{P}$ by $M_\ast^{2+n} \sim M_P^2 R^{-n}$,
where $R$ and $n$ are the size and number of extra dimensions respectively.
This means that $M_{\ast }$ may be as small as ${\mathcal {O}}$(TeV),
raising the exciting prospect of observing quantum gravity effects
in the near future
(see the reviews \cite{Kanti,reviews,Harris} for details
and references).

Amongst quantum gravity processes, the potential to create
higher-dimensional black holes in
trans-Planckian particle collisions at the LHC has
attracted intense interest in the literature.
It is anticipated that such mini black holes will evaporate very quickly,
due to Hawking radiation \cite{hawking}.
Theoretically, the evaporation of higher-dimensional mini black holes is modelled
in terms of four stages:
the so-called `balding', `spin-down', `Schwarzschild', and
`Planck' phases \cite{giddings}.
Of these, the `Schwarzschild' phase is the simplest to study and
the Hawking emission during this phase has been analyzed in depth
by a number of authors, using both analytical
\cite{kmr1, Frolov1} and numerical \cite{HK1, graviton-schw}
techniques.

The `spin-down' phase of the evolution has received considerable attention recently
\cite{DHKW, CKW, CDKW1, IOP, rot-other,flachi}.
A complete analysis of the `spin-down' phase is lacking because
the graviton emission has yet to be fully modelled (see \cite{graviton-rot} for
work in this direction), leaving open the question of whether the idea that
``black holes radiate mainly on the brane'' \cite{emparan} is valid
for this phase of the evolution
(see \cite{HK1, CDKW1, brane-bulk} for some papers discussing this issue).

The effect of including the `spin-down' phase in simulations of black hole
events at the LHC has recently been examined in detail \cite{charybdis2}
%(for an earlier work, see \cite{blackmax}).
(having been originally considered in \cite{giddings}, and also in \cite{blackmax}).
It is found that, even if the black hole loses a substantial amount of
angular momentum during the `balding' phase, the `spin-down' phase makes
a significant difference to the properties of black hole events, which cannot be
adequately approximated by the isotropic evaporation of the `Schwarzschild' phase.
Another key finding of \cite{charybdis2} is that the `spin-down' phase is not,
as previously thought, short-lived compared with the `Schwarzschild' phase
(in fact, it is not possible to fully distinguish a separate `Schwarzschild'
phase, with the rotation of the black hole being significant for most of
its lifetime).

The significance of the `spin-down' phase complicates black hole events at the LHC
\cite{charybdis2}, making their detection more challenging.
It also means that
measuring all the parameters describing a particular black hole event
is likely to be very difficult.
Our purpose in this letter is to investigate
a key difference in the evaporation during the `spin-down'
phase compared with the `Schwarzschild' phase, namely the non-isotropy of the emission.
We are seeking aspects of the anisotropic emission which are
particularly sensitive to the parameters describing the higher-dimensional black hole,
with the intention that this will aid the development of
detection and measurement strategies using full Monte-Carlo simulations
\cite{charybdis2}.

The structure of this letter is as follows.
In section \ref{sec-theory}, we briefly review the mathematical description of the
higher-dimensional rotating black hole and Hawking radiation from it.
In section \ref{sec-profile}
we consider the emission, on the brane, of positive and negative helicity
spin-1/2 and spin-1 particles.
We focus on the angular distribution of the emitted energy by these particles,
and how this depends on the energies of the individual emitted
particles, the angular momentum of the black hole, and the number of extra
dimensions.
Finally, in section \ref{sec-conclusions}, we present our conclusions,
making some remarks on the prospects of using our results
to determine some of the properties of rotating higher-dimensional black holes
at the LHC.

%%%%%%%%%%%%%%%%%%%%%%%%%%%%%%%%%%%%%%%%%%%%%%%%%%%%%%%%%%%%%%%%%%%%%%%

\section{Theoretical background and field equations
\label{sec-theory}}

The gravitational field around a ($4+n$)-dimensional uncharged
rotating black hole is described by the well-known Myers-Perry solution \cite{MP}.
The projected line-element follows by fixing the values of the additional angular
coordinates describing the extra space-like dimensions \cite{Kanti}. Then, the
gravitational field on the brane takes the form
\begin{eqnarray}
ds^2 & =& -\left( 1 - \frac{\mu}{\Sigma r^{n-1}} \right) dt^2
- \frac{2 a \mu \sin^2 \theta}
{\Sigma r^{n-1}}\,dt \, d\varphi
%\nonumber \\ & &
 + \left( r^2 + a^2 + \frac{a^2 \mu \sin^2 \theta}{\Sigma r^{n-1}} \right) \sin^2
\theta\,d \varphi^2
\nonumber \\ & &
+ \frac{\Sigma}{\Delta}\,dr^2
+ \Sigma\,d\theta^2
\,,
\label{metric}
\end{eqnarray}
where
\begin{equation}
\label{sigma}
\Delta = r^2 + a^2 - \frac{\mu}{r^{n-1}}, \quad \quad
\Sigma = r^2 + a^2 \cos^2 \theta\,.
\end{equation}
We have also assumed that the black hole metric has only one
non-zero angular momentum
component, in a plane parallel to the brane, as we are interested
in black holes created
by the collision of particles on the brane.
The mass  $M_{BH}$ of the black hole and its
angular momentum $J$ are then proportional to $\mu$ and $a\mu$, respectively,
\begin{equation}
M_{BH} = \frac{(n+2) A_{n+2}}{16 \pi G}\,\mu , \quad \quad
J = \frac{2}{n+2} M_{BH}\,a\,,
\end{equation}
where $A_{n+2} = 2\pi^{(n+3)/2} / \Gamma[(n+3)/2]$ is the area of
an $(n+2)$-dimensional
unit sphere, and $G$ is the $(4+n)$-dimensional Newton's constant.
By demanding that
the impact parameter
between the colliding particles is small enough for a black hole to be created,
an upper bound can be imposed on the angular momentum parameter $a$ as follows:
we have $a_* \leq a_*^{max}=n/2+1$ \cite{Harris}, where $a_*=a/r_h$. The
radius of the black hole's event horizon $r_h$ is the largest, positive root  of
$\Delta(r) = 0$, and, for $n \ge  1$, there is only one such root in the region
$r > 0$,
which may be implicitly written as $r_h^{n+1}=\mu/(1+a_*^2)$.

The four-dimensional background (\ref{metric}) is the one felt by the brane-localized
Standard Model fields and thus the one that should be used for the derivation
of the field
equations for scalars, fermions and gauge bosons. By using the Newman-Penrose
formalism
and generalizing Teukolsky's four-dimensional analysis \cite{Teukolsky}, one may
derive a
`master' partial differential equation for the field perturbation
$\Psi_h$
on the brane \cite{Kanti, DHKW, CKW}.
Here, $h$ is the helicity, or spin-weight,
$h=(-|s|,+|s|)$, that we use to distinguish the radiative components of
the spin-$s$ field.
The brane `master' equation for the particular background (\ref{metric})
turns out to be separable - by
using a factorized ansatz for the field perturbation expressed as a Fourier
mode series
%%%%%%%%%
\begin{equation}
\Psi_h(t,r,\theta, \varphi)  =
%\int_{-\infty}^{+\infty} d\omega\,\sum_{l=|h|}^{+\infty}
%\,\sum_{m=-l}^{+l}
\sum _{\Lambda }
{}_h a_\Lambda \,_hR_\Lambda(r)\,_hS_\Lambda(\theta)\,e^{-i\omega t}\,
e^{im\varphi}\,,
\label{fact}
\end{equation}
%%%%%%%%%%
where $_ha_\Lambda$ are the Fourier coefficients,
we write $\Lambda=\{lm\omega\}$ to denote the
set of `quantum numbers' of each mode, and $_hS_\Lambda$ are the spin-weighted
spheroidal harmonics, the `master' equation leads to two decoupled ordinary differential
equations, namely
%%%%%%%%%%%
\begin{equation}
0  =
\Delta^{-h}\frac{d\,}{dr}\left(\Delta^{h+1} \frac{d_hR_\Lambda}{dr}\right) +
\left[\frac{K^2-ihK \Delta'(r)}{\Delta}
%\right. \nonumber \\ & & \left.
+4ih\omega r + h(\Delta''(r)-2) \delta_{h,|h|} -
{}_h\lambda_\Lambda\right] {}_hR_\Lambda, \label{radial}
\end{equation}
and
%%%%%%%%%
\begin{equation}
0  =
\left[\frac{d\,}{dx} \left((1-x^2) \frac{d\,}{dx}\right)
+a^2 \omega^2(x^2-1) +2ma\omega
% \right. \nonumber \\ & & \left.
-
2hma\omega x-
\frac{(m+hx)^2}{1-x^2} +{}_h\lambda_\Lambda
+h\right] {}_hS_\Lambda(x) \,.
%\nonumber \\
\label{angular}
\end{equation}
%%%%%%%%%%%%%%%%
In the above, we have defined  the quantities $K=(r^2+a^2) \omega -am$
and $x=\cos \theta$, and $_h\lambda_\Lambda$ is the constant of separation
between the radial and angular equations.

The black-hole background (\ref{metric})
emits elementary particles on the brane in the form of Hawking radiation
with temperature
%%%%%%%%%%%%%%%
\begin{equation}
T_H=\frac{(n+1) + (n-1) a_*^2}{4\pi\,(1+a_*^2)\,r_h} .
\label{T_H}
\end{equation}
%%%%%%%%%%%%%
In this letter, we focus on the power emission rate,
that may be expressed in two forms: in terms of unit time and energy
and integrated over all angles $\theta$
%%%%%%%%%%%%%
\begin{equation}
\frac{d^2E}{dt\,d\omega}=\frac{1+\delta_{|s|,1}}{2\pi}\,
\sum_{l=|s|}^{\infty}\,\sum_{m=-l}^{+l}\,\frac{\omega}{\exp(\tilde \omega/T_H) \pm 1}\,
{\bf T}_\Lambda\,,
\label{power-int}
\end{equation}
%%%%%%%%%%%%%%%
or in terms of unit time, frequency and angle of emission,
%%%%%%%%%%%%%
\begin{equation}
\frac{d^3E}{d(\cos\theta)\,dt\,d\omega}  =
\frac{1+\delta_{|s|,1}}{4\pi}\,
\sum _{l,m}
%\sum_{l=|s|}^{\infty}\,\sum_{m=-l}^{+l}\,
\frac{\omega}{\exp(\tilde \omega/T_H) \pm 1}
%\nonumber \\ & & \qquad \times
{\bf T}_\Lambda\,\left({}_{-h}S^2_\Lambda+{}_hS^2_\Lambda\right)\,.
\label{power-ang}
\end{equation}
%%%%%%%%%%%%%%%
In the above expressions, $\tilde \omega=\omega-am/(a^2+r_h^2)$
and $\pm 1$ is a statistics
factor for fermions and bosons, respectively. The extra factor $\delta_{|s|,1}$ in
the emission rates for gauge bosons comes from the two linearly independent polarization
states of the spin-1 field. Finally, the quantity ${\bf T}_\Lambda$ is the transmission
probability, defined as the flux of energy transmitted down the event horizon over the
incident flux of energy on the black hole -- it is this quantity that modifies the
form of the radiation spectra of a black hole from the purely thermal one of
a black body.
For details of the derivation and computation of the emission
rates (\ref{power-int}, \ref{power-ang}),
including the numerical techniques used in solving the differential equations
(\ref{radial}, \ref{angular}),
we refer the reader to Refs.~\cite{DHKW, CKW}.

In Refs.~\cite{DHKW, CKW}, the emission rates for
scalars, fermions and gauge bosons were
studied in detail.
Amongst the results in those papers, it was found that
the emitted power was significantly enhanced with both $n$ and $a_*$
- the enhancement factor
for all species of particles was of order ${\cal O}(100)$ as $n$ changed from 1 to 6,
and of order ${\cal O}(10)$ as $a_*$ increased from zero
towards its maximum value $a_*^{max}$.
The differential emission rate per unit time, frequency and angle of emission
(\ref{power-ang}) was also studied.
The angular variation in the spectra was characterized by two main features:
in the low-energy regime, the
spin-rotation coupling for fermions and gauge bosons caused the polarization
of the emitted radiation
in directions nearly parallel to
the rotation axis; as either the angular momentum of
the black hole and/or the energy of the emitted particle increased, this feature
receded and the centrifugal force caused the concentration of the emitted radiation,
for all species of particles, on the equatorial plane, i.e. transversely to the
rotation axis of the black hole.

%%%%%%%%%%%%%%%%%%%%%%%%%%%%%%%%%%%%%%%%%%%%%%%%%%%%%%%%%%%%%%%%%%%%%%%%%%%%%%%%%

\section{The angular profile
\label{sec-profile}}

In this section, we will investigate further the angular profile of the emitted
radiation on our brane from a higher-dimensional black hole by focusing on the
emission of individual helicity modes. We will therefore consider only the
emission of brane-localized fermions and gauge bosons, and ignore henceforth
the single-component scalar fields.
In each case, the energy flux for positive helicity particles is obtained from
(\ref{power-ang}) by including only the term with ${}_{-|s|}S^2_\Lambda$;
similarly the energy flux for negative helicity particles is obtained
from the ${}_{+|s|}S^2_\Lambda$ term in (\ref{power-ang}) \cite{helicity}.

The study of the differential emission rate (\ref{power-ang}) for
spin-1/2 and spin-1 particles of Ref.~\cite{CKW} considered only
the `total' emission, i.e. the sum of the emission of both
helicity modes in each case. The spin-weighted spheroidal
harmonics satisfy the property (after an overall choice of sign)
%%%%%%%%%%%%
\begin{equation}
{}_hS_\Lambda(\theta)=(-1)^{l+m}\,{}_{-h}S_\Lambda(\pi-\theta)\,.
\label{Sproperty}
\end{equation}
%%%%%%%%%%%
As a result, the sum of the squared values of the spin-weighted spheroidal
harmonics for the two helicities $h=(-|s|,|s|)$ is bound to be symmetric under
the change $\pi \rightarrow \pi-\theta$,
and the plots showing the angular variation in the `total' emission are
symmetric in the two hemispheres. However, it is clear that the emission
of single-helicity modes is not going to have the same property,
and it is this additional variation in the angular distribution that we
investigate here. With a similar motivation, certain aspects of the
angular distribution of the emission spectra of spin-1/2 particles were
studied also in \cite{flachi}.

%%%%%%%%%%%%%%%%%%%%%%
\begin{figure}[t]
\begin{center}
\begin{tabular}{lll}
Fermions & &
\\ \hspace*{-0.4cm}
\includegraphics[width=5.8cm]{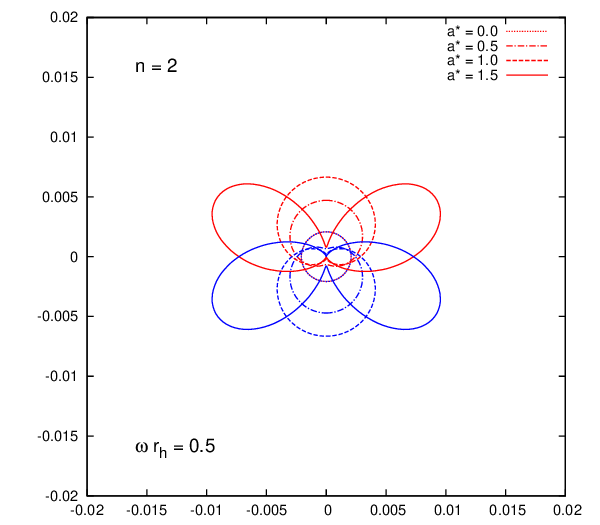}
& \hspace*{-0.9cm}
\includegraphics[width=5.8cm]{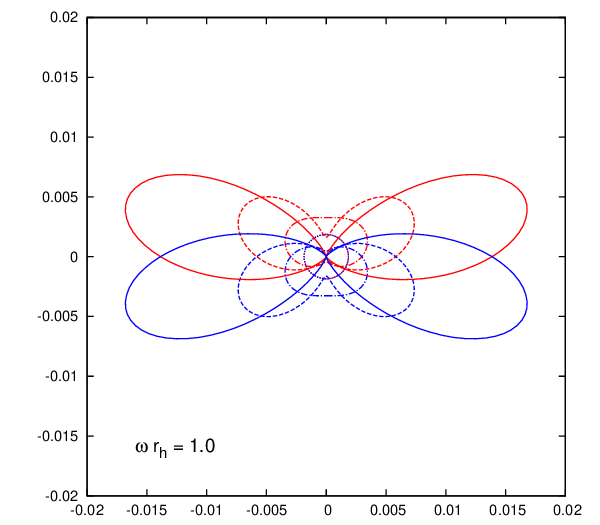}
& \hspace*{-0.9cm}
\includegraphics[width=5.8cm]{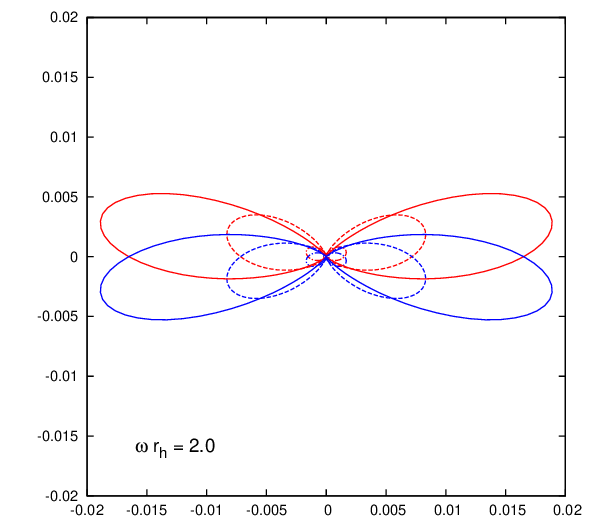}
\\
Vector Bosons & &
\\ \hspace*{-0.4cm}
\includegraphics[width=5.8cm]{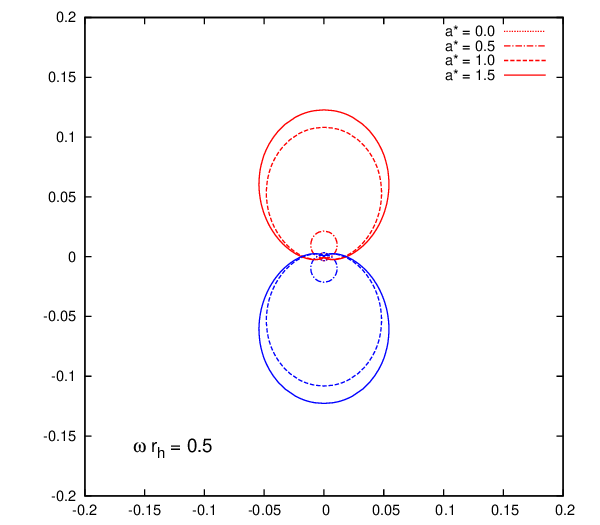}
& \hspace*{-0.9cm}
\includegraphics[width=5.8cm]{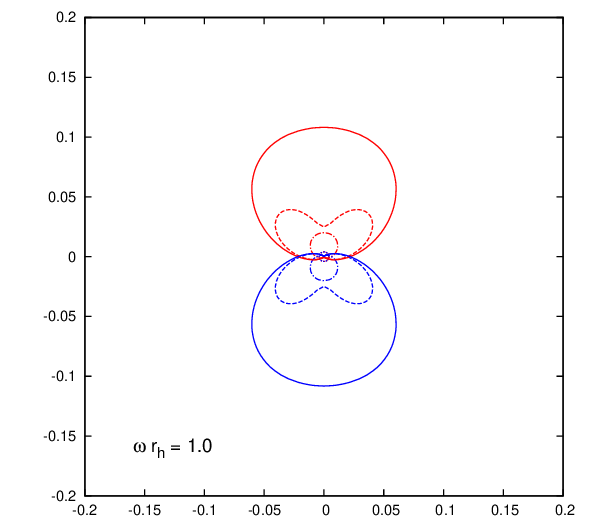}
& \hspace*{-0.9cm}
\includegraphics[width=5.8cm]{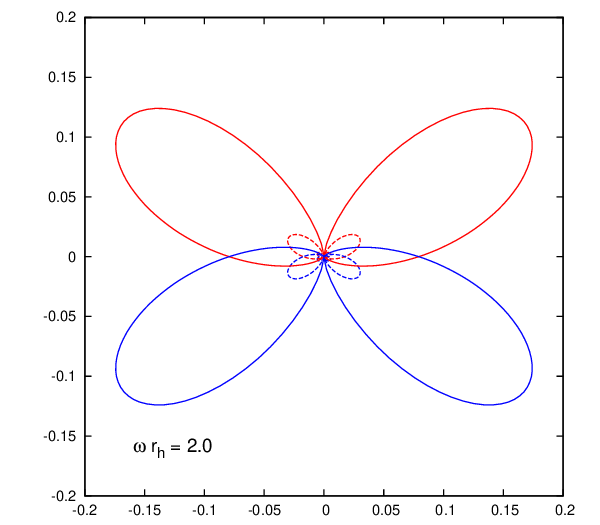}
\end{tabular}
\end{center}
\caption{Polar plots depicting the emitted
differential power rate (\ref{power-ang}) of individual helicities for
brane localized fermions (top) and vector bosons (bottom) in terms of the angle.
The red curves denote positive helicity emission
and the blue curves negative helicity.
The number of extra dimensions is fixed by $n=2$, and we have plotted the
angular distribution for different $\omega r_{h}$, and
for four values of the angular
momentum parameter $a_{*}$.
In each case the distance of a point on a curve from the origin is the
magnitude of the power emitted in that direction.}
\label{polar}
\end{figure}
%%%%%%%%%%%%%%%%%%%%%%

In Fig.~\ref{polar}, we present a set of polar plots for the differential power
emission rate (\ref{power-ang})
for fermions (top) and gauge bosons (bottom)
in terms of the emitted latitudinal angle $\theta $.
The two colour schemes
represent the two opposite helicities $h>0$ (red) and $h<0$ (blue).
The three plots for each type of particle
correspond to different values of the energy channel ($\omega r_h=0.5, 1.0, 2.0$),
 four values of the angular-momentum parameter of the black hole
($a_*=0.0, 0.5, 1.0$ and $1.5$)
and fixed number of additional space-like dimensions ($n=2$). The rotation axis
in these plots runs vertically.
For each curve in the graphs in Fig.~\ref{polar},
the distance of a point on a curve from the origin is the
magnitude of the power emitted in that direction.
For each value of the emitted particle energy,
the emission in the zero-angular-momentum
case $a_{*}=0$ is isotropic as expected,
and represented by a small circle in the centre of the plots.

From Fig.~\ref{polar}, we observe that
for low values of the energy of the emitted particle
and small black hole angular momentum $a_{*}$, the two helicity modes are
indeed aligned to the rotation axis of the black hole, in accordance to the behaviour
found in \cite{CKW}, but with the two helicity `jets' facing towards opposite
sides.
Positive helicity particles
are emitted in the `northern' hemisphere (i.e. pointing in the same direction
as the angular velocity vector of the black hole),
while negative helicity particles
are emitted in the `southern' hemisphere.
We also note the order of magnitude increase in the emitted power in
spin-1 particles compared with spin-1/2 particles.

As the energy of the emitted particles is increased,
or if the black hole angular momentum $a_{*}$ is very large,
this pattern is
destroyed: although the preference shown by each helicity mode towards one of the
hemispheres is still prominent, the emission is now concentrated in two directions
at almost 45 degrees to the rotation axis.
This effect is stronger for fermions than for the gauge bosons.
The energy at which this effect becomes apparent increases as the angular momentum
of the black hole decreases: for large $a_{*}=1.5$, the `jets' at 45 degrees have formed
even when $\omega r_{h}$ is as low as $0.5$, while for small $a_{*}=0.5$,
we see these `jets' at 45 degrees only when $\omega r_{h}$ is nearer 2.
As the energy increases
further, the asymmetry in the emission between the two helicity modes becomes gradually
less prominent with the emission concentrating more and more on the equatorial plane.

It can also be seen from Fig.~\ref{polar} that increasing $a_{*}$ increases
the magnitude of the power emitted,
particularly for higher energy modes (and
hence increases the number of particles emitted, as observed in \cite{flachi}).
There is also a certain degeneracy in the shape of the curves, for example,
the curves for fermion emission with $\left( a_{*}=1.5, \omega r_{h}=0.5 \right)$ and
$\left( a_{*} = 1.0, \omega r_{h} = 1\right) $ are very similar.
However, since the energy of the emitted particles can be measured experimentally,
this degeneracy can be broken.

%%%%%%%%%%%%%%%%%%%%%%
\begin{figure}[t]
\begin{center}
\begin{tabular}{lll}
Fermions & &
\\ \hspace*{-0.5cm}
\includegraphics[width=5.8cm]{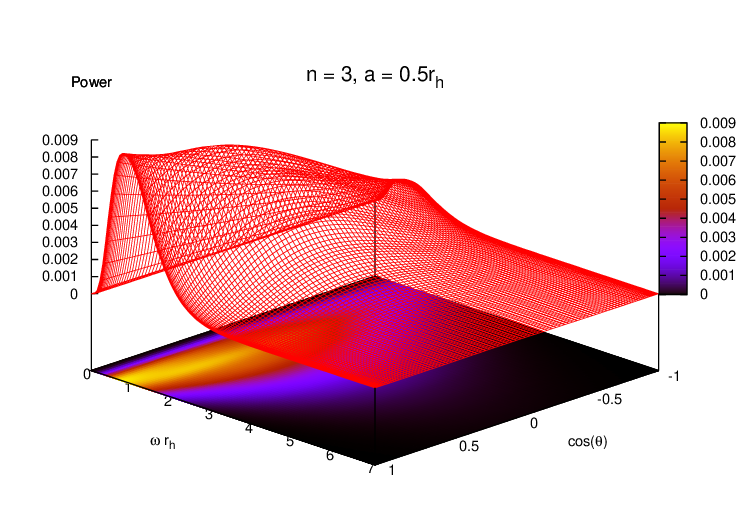}
& \hspace*{-0.9cm}
\includegraphics[width=5.8cm]{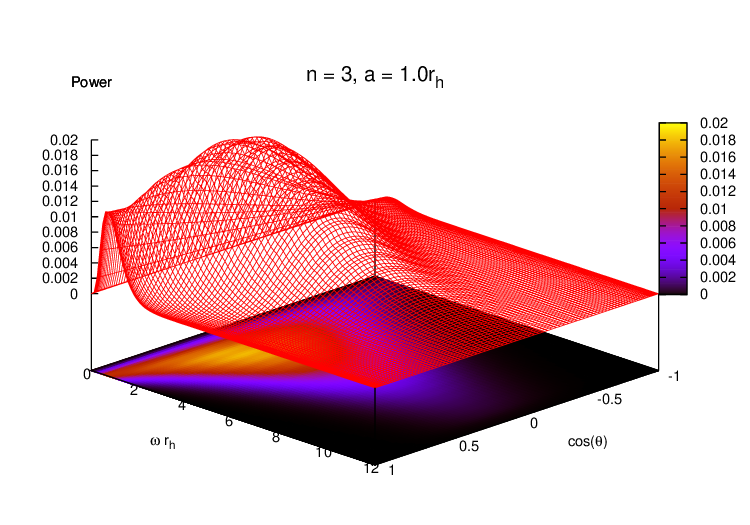}
& \hspace*{-0.9cm}
\includegraphics[width=5.8cm]{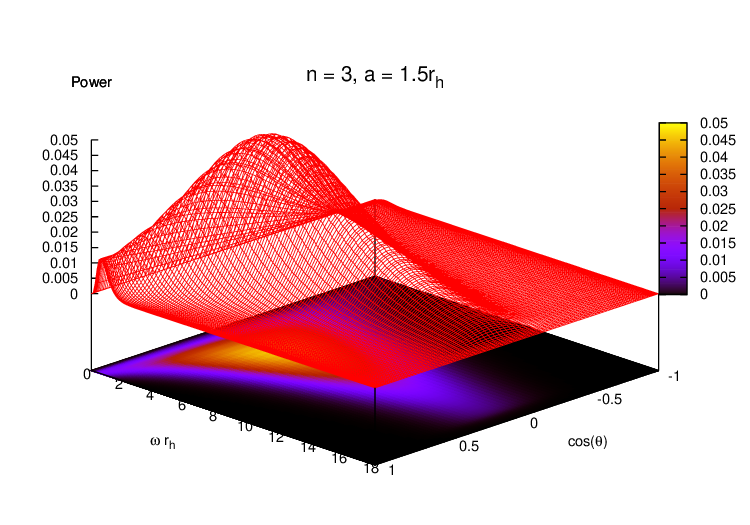}
\\
Vector Bosons & &
\\ \hspace*{-0.5cm}
\includegraphics[width=5.8cm]{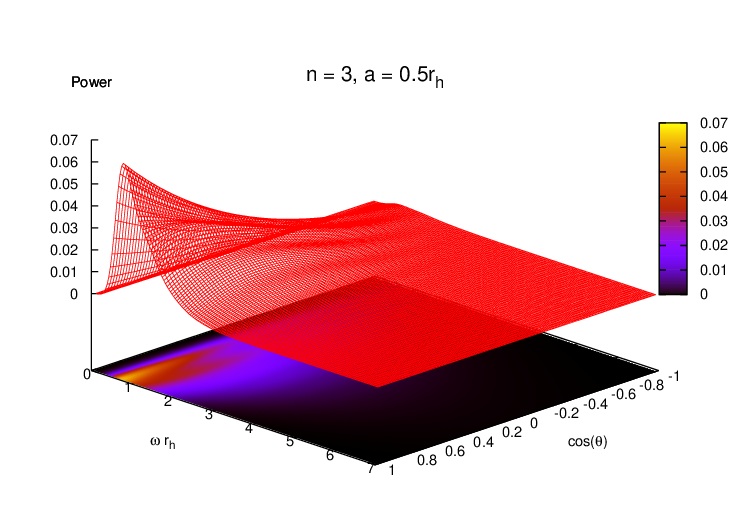}
& \hspace*{-0.9cm}
\includegraphics[width=5.8cm]{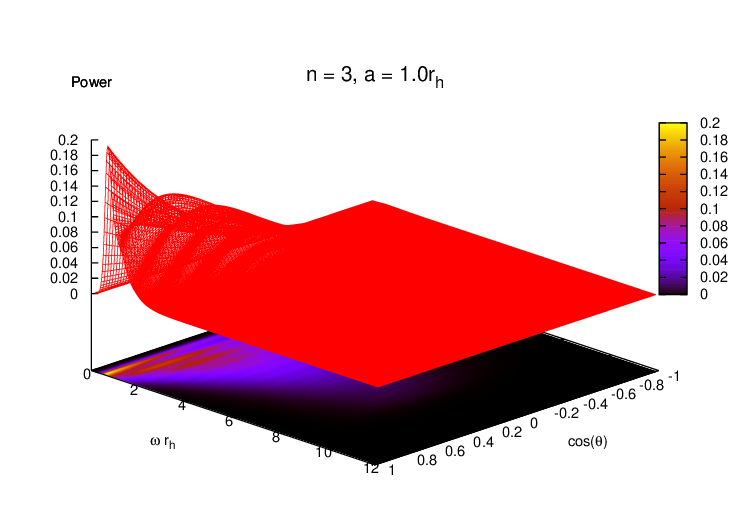}
& \hspace*{-0.9cm}
\includegraphics[width=5.8cm]{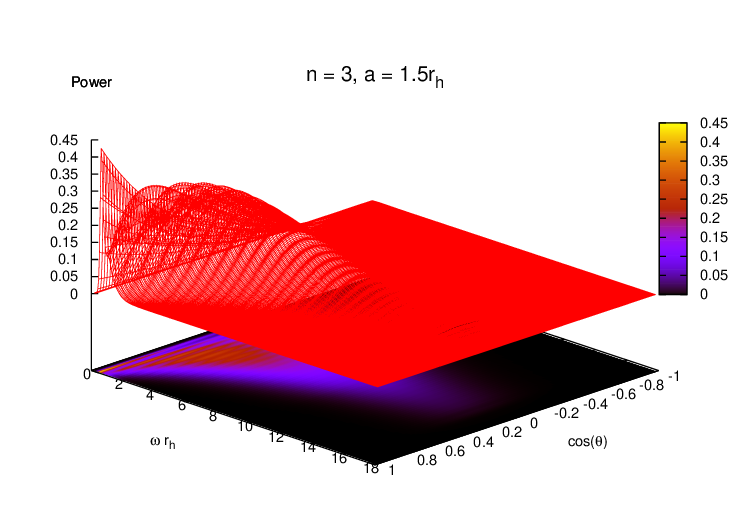}
\end{tabular}
\end{center}
\caption{Power-versus-frequency-versus-angle plots for
positive helicity fermion (top) and vector boson (bottom) emission,
for $n=3$, and $a_*=0.5, 1.0, 1.5$.}
\label{3D}
\end{figure}
%%%%%%%%%%%%%%%%%%%%%%

To explore the dependence of the angular distribution on the particle energy further,
in Fig.~\ref{3D}, we have plotted the emitted power for positive
helicity fermions (top) and gauge bosons (bottom)
as a function of the frequency $\omega r_{h}$ and $\cos \theta $,
considering three different values for the angular-momentum
parameter ($a_*=0.5, 1.0, 1.5$) and fixed dimensionality of space-time ($n=3$).
For smaller non-zero values of
$a_*$, almost all of the radiation of the individual helicity mode is emitted
in directions nearly parallel to
the rotation axis of the black hole with a clear preference towards one
of the hemispheres; the effect is more marked
for spin-1 particles, as might be expected from the increased spin-orbit coupling due
to the increased internal spin of the vector bosons compared with the fermions.
As $a_*$ increases, the power emission curve for fermions becomes higher,
as expected, and also more and more radiation is concentrated on
the equatorial plane
rather than in directions nearly parallel to the rotation axis.
However, for spin-1 particles, we can see in Fig.~\ref{3D}
that, while for vector bosons emitted with higher energies there
is evidence for the emission tending to be nearer the equatorial plane than
the axis, the majority of the emission in lower-energy particles is still
very much concentrated in directions nearly parallel to
the axis of rotation.

Therefore, the emission of lower-energy bosons
in directions nearly parallel to the axis of rotation may
provide the best means of measuring the direction of the axis of rotation.
On the other hand, if the energy of the emitted fermions is accurately measured,
the proportion of the emitted radiation
in directions nearly parallel to the rotation axis and close to the
equatorial plane (whose directions can, at least in theory, be determined
from the vector boson emission) can give us an estimate for the angular-momentum
of the black hole  (see also \cite{flachi} for a discussion of this measurement).

%%%%%%%%%%%%%%%%%%%%%%
\begin{figure}[t]
\begin{center}
\begin{tabular}{lll}
Fermions & &
\\ \hspace*{-0.4cm}
\includegraphics[width=5.6cm]{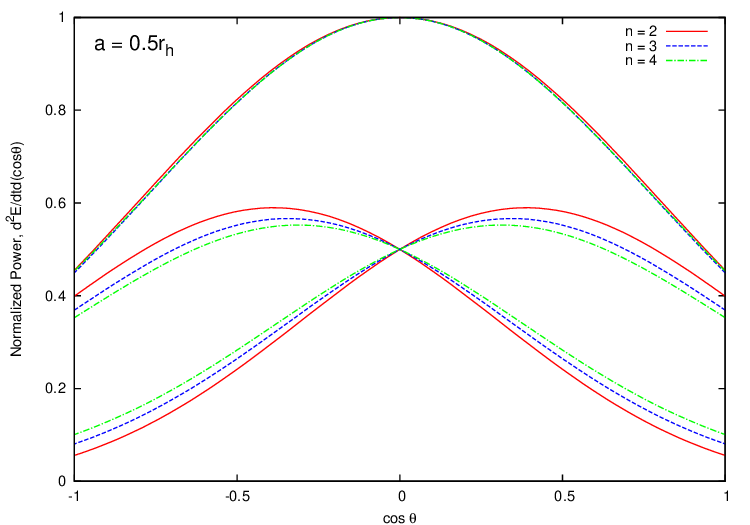}
& \hspace*{-0.7cm}
\includegraphics[width=5.6cm]{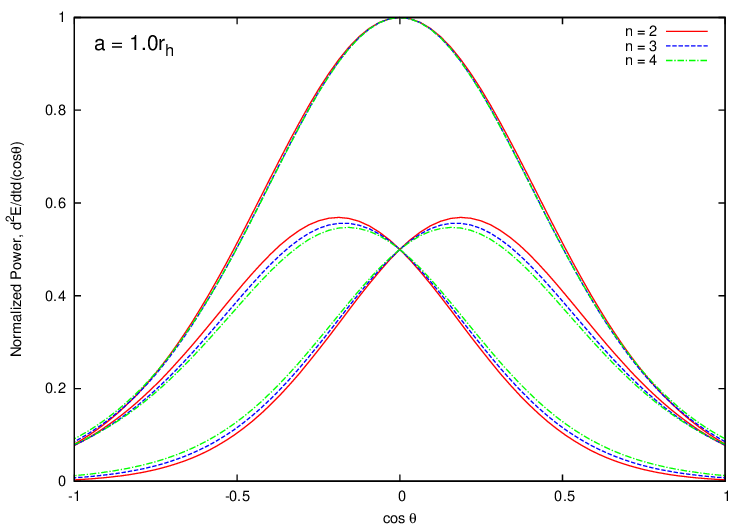}
& \hspace*{-0.7cm}
\includegraphics[width=5.6cm]{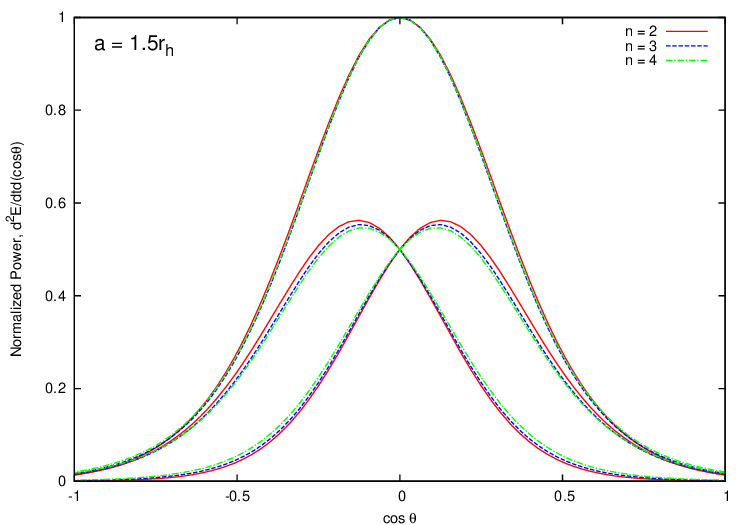}
\\
Vector Bosons & &
\\ \hspace*{-0.4cm}
\includegraphics[width=5.6cm]{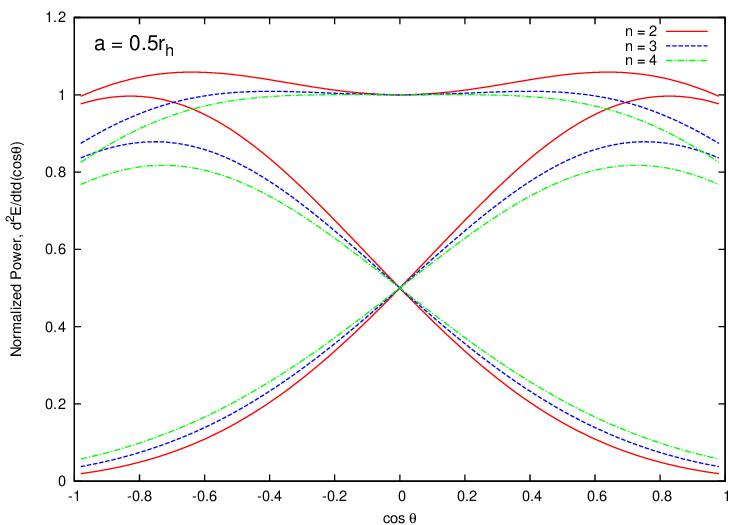}
& \hspace*{-0.6cm}
\includegraphics[width=5.6cm]{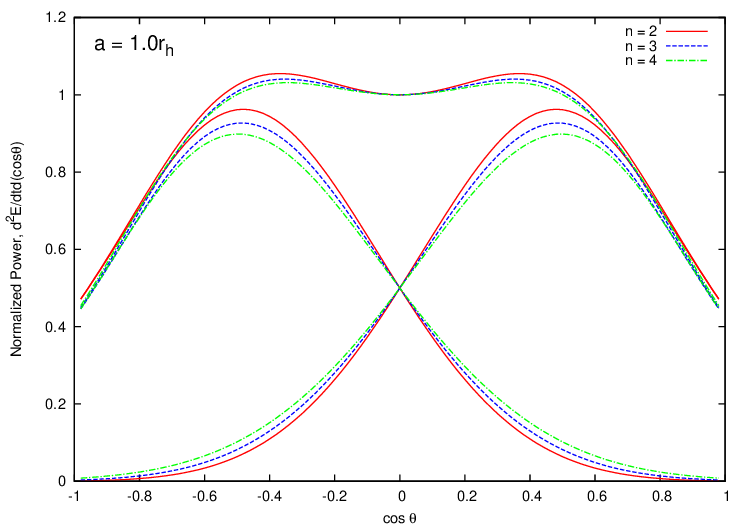}
& \hspace*{-0.6cm}
\includegraphics[width=5.6cm]{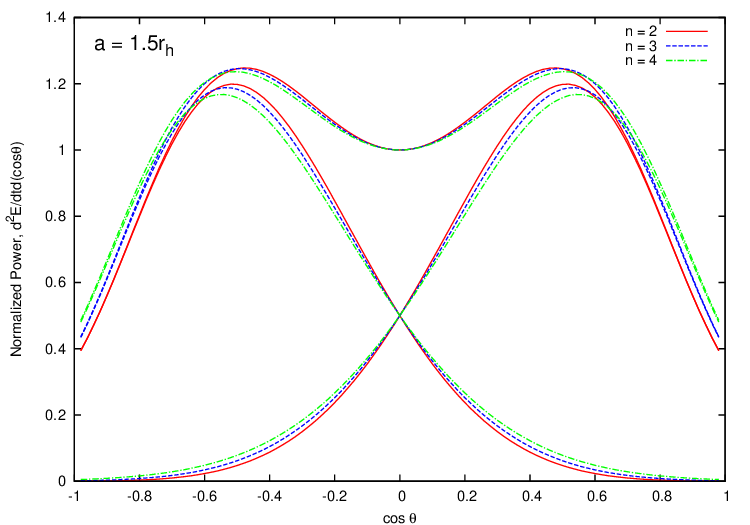}
\end{tabular}
\end{center}
\caption{Energy emissivity integrated over frequency $\omega r_{h}$,
for fermions (top) and
vector bosons (bottom) as a function of
angle, for $n=2,3,4$, and $a_*=0.5, 1.0, 1.5$.
The
power has been normalized so that the total emission when $\theta = \pi /2$
is equal to unity.
In each plot, the upper
curve corresponds to the emission rate of the sum of both helicities and
the two lower ones to the ones of the two individual helicities.}
\label{n}
\end{figure}
%%%%%%%%%%%%%%%%%%%%%%

Thus far we have not considered the role the number of extra
dimensions plays in this angular distribution. In the case of the
`total' emission rates, where the contribution from both helicity modes was
considered, the dependence on $a_*$ and $n$ was also very prominent, however,
as both parameters caused the same effect in the spectra, namely
their enhancement, it was impossible to distinguish their role.
In Fig.~\ref{n}, we present a set of plots
of the energy emissivity of the black hole integrated over the
frequency
$\omega r_h$, in terms of the emitted angle for three different values of
the angular momentum parameter ($a_*=0.5, 1.0, 1.5$), and for three different
values of the number of extra dimensions ($n=2,3,4$) in each case. The purpose
is to investigate the dependence of the {\em {shape}} of the curves on $n$,
rather than the magnitude of the emission.
To this end, we rescale our data so that the peak
of the emission curve for the sum of the two helicity modes is at unity at
$\theta=\pi/2$ with the rescaled integrated emissivity curves for each helicity mode shown
below.

The spin-1 profiles in Fig.~\ref{n} vary more than the
spin-1/2 profiles as the number of
extra dimensions, $n$, changes, but none of the profiles change a great deal except
in the spin-1 case
when the angular momentum of the black hole is comparatively small ($a_{*}=0.5$).
From
this we conclude that the angular distribution of emitted
particles is not a good indicator of the number of extra dimensions. However,
this also means that the angle-dependent spectra carry a strong dependence
only on the angular momentum parameter, and this opens the way for the
determination of $a_*$.
From Fig.~\ref{n}, we see that
the profiles of the individual helicity particles are more sharply peaked
for spin-1 compared with spin-1/2 particles, at the same value of $a_{*}$,
but, interestingly, the total power emission (summing over the two helicities)
has a broader shape in the spin-1 case compared with the spin-1/2 case,
for all values of $a_{*}$.

The authors of \cite{flachi} also studied the angular distribution of the flux of
negative helicity fermions, but for a particular value of the frequency $\omega r_{h}$.
Fig.~\ref{3D} confirms, for all frequencies in the power
flux, their observation for the fermion particle flux that there is an asymmetry in the
emission in helicity-dependent states, which becomes more strongly peaked as the
angular-momentum parameter $a_{*}$ increases.
From Fig.~\ref{n}, it can also be seen that the asymmetry in the power
flux decreases as the number of extra dimensions $n$ increases, which
was found to be the case for the fermion particle flux at a particular frequency in
\cite{flachi}.

\section{Discussion and conclusions
\label{sec-conclusions}}

The possibility of creating higher-dimensional black holes at the LHC
and subsequently observing their Hawking radiation is one of the most
exciting consequences of models with Large Extra Dimensions.
The recent inclusion of black hole angular momentum in simulations of such black hole
events \cite{charybdis2,blackmax}, as well as improving the accuracy of
the simulations, has also indicated that black hole rotation is likely
to have major consequences for the detection of Hawking radiation.
This is particularly true for the emission, on the brane, of particles
with non-zero spin, which, in practice, are the main particles which will be
detected should black holes be formed at the LHC.

The primary consequence of including black hole rotation is that the Hawking emission
is no longer isotropic \cite{giddings, CKW, IOP, flachi}.
In this letter we have examined further the nature of this anisotropy for particles of
spin-1/2 and spin-1, considering separately the positive and negative helicity states,
to determine whether any features of the anisotropic emission could,
at least in principle, be used to derive information about the evaporating
black hole or the higher-dimensional space-time.

There are three main space-time parameters which it would be desirable to
probe in black hole events: the number of extra dimensions $n$,
the size of the extra dimensions $R$, and the
energy scale of quantum gravity $M_{\ast }$.
These will be the same for all black hole events, so a large number of events
can be used in their determination.
In addition, one would be interested, for individual black hole events,
in measuring the mass $M$ and angular momentum $J$
of the black hole when it is first formed
(or, equivalently, $r_{h}$ and $a_{*}$).

The question of how to extract information about these parameters
from the Hawking radiation is still
not completely resolved.
For rotating black holes, since increasing either $n$ or $a_{*}$ increases
the Hawking flux,
disentangling this degeneracy in the values of $n$ and $a_{*}$ from, for example,
the numbers of emitted particles, may be difficult.
In this letter, we have found that, for both fermion and gauge boson emission,
the shape of the angular distribution of total energy emission
for both positive and negative helicity particles depends only weakly on the
number of extra dimensions $n$, for all values of the
black hole angular momentum parameter $a_{*}$.
This means that the shape cannot be used to measure the value of $n$,
with only the magnitude of the power emitted depending strongly on its
value. On the other hand, the angular distribution of energy emission
for both species of particles was found to depend strongly on $a_{*}$
and can therefore be used as a good indicator of the angular momentum
of the produced black hole \cite{giddings}.
By measuring $a_{*}$ for many black hole events, it would be possible to
build up a distribution of $a_{*}<a_*^{max}$.

Another interesting measurement would be the direction of the axis of
rotation of the black hole \cite{giddings}.
Rotating black holes preferentially emit spin-1 particles over fermions
(with approximately an order-of-magnitude greater energy emission)
and low-energy spin-1 particles are emitted almost entirely
in directions nearly parallel to the axis
of rotation, particularly for smaller values of $n$.
On the other hand, the fermion energy emission is strongly peaked on the
equatorial plane, providing complementary evidence for the direction of the
axis of rotation \cite{flachi}.
Measuring particles of a specific helicity further refines this measurement
by effectively indicating the `arrow' of the black hole angular velocity vector
along this axis.

In our study of the physics of Hawking emission from higher-dimensional
black holes, we have taken the usual semi-classical approach of
calculating the emission
from a fixed black hole geometry.
This does not model the evolution of the black hole during the evaporation
process, with the black hole losing mass and angular momentum.
It is also anticipated that the black hole will recoil as it emits particles
\cite{Frolov1},
due to momentum conservation, and that the axis of rotation may well change
direction.
Every particle emitted, observed or unobserved, changes the mass and angular
velocity of the black hole.
In practice, the process is stochastic and hence each black hole decay is unique.
Nevertheless, by combining the spectra of semi-classical theory with Monte Carlo
methods (as in \cite{charybdis2,blackmax}), one may model the likely `ensemble'
of black hole decays.
This approach will be crucial for determining whether the physics
effects we have described in this letter will be detectable in real data
and for devising detection strategies.

\section*{Acknowledgements}
We thank V.~P.~Frolov for useful discussions on this topic.
MC and SRD gratefully acknowledge financial support from the Irish
Research Council for Science, Engineering and Technology (IRCSET).
The work of MC is partially funded by Funda\c{c}\~{a}o para a Ci\^{e}ncia e
Tecnologia (FCT) - Portugal, reference PTDC/FIS/64175/2006.
PK acknowledges participation in
RTN Universenet (MRTN-CT-2006035863-1 and MRTN-CT-2004-503369).
EW thanks the School of Mathematical Sciences, University College Dublin,
for hospitality while this work was completed.
The work of EW is supported by STFC (UK), grant number ST/G000611/1.

\end{document}